\documentclass[12pt]{article}
\usepackage{amsmath,amssymb,graphicx,color}

\newwrite\ffile\global\newcount\figno \global\figno=1

\def\writedef#1{}

\input epsf
\def\figin{\epsfcheck\figin}\def\figins{\epsfcheck\figins}
\def\epsfcheck{\ifx\epsfbox\UnDeFiNeD
\message{(NO epsf.tex, FIGURES WILL BE IGNORED)}
\gdef\figin##1{\vskip2in}\gdef\figins##1{\hskip.5in}
\else\message{(FIGURES WILL BE INCLUDED)}%
\gdef\figin##1{##1}\gdef\figins##1{##1}\fi}

\def\figinsert{}
\def\ifig#1#2#3{\xdef#1{fig.~\the\figno}
\writedef{#1\leftbracket fig.\noexpand~\the\figno}%
\figinsert\figin{\centerline{#3}}\medskip\centerline{\vbox{\baselineskip12pt
\advance\hsize by -1truein\center\footnotesize{  Fig.~\the\figno.}
#2}}
\bigskip\endinsert\global\advance\figno by1}
\def\endinsert{}

\usepackage{amssymb}
\usepackage{graphics}
\begin{document}
\baselineskip 18pt \providecommand{\SU}[1]{}
\renewcommand{\SU}[1]{\ensuremath{\mathrm{SU}(#1)}}
\providecommand{\UI}{}
\renewcommand{\UI}{\ensuremath{\mathrm{U}(1)}}
\providecommand{\psgroup}{}
\renewcommand{\psgroup}{\ensuremath{\mathrm{SU}(4) \otimes \mathrm{SU}(2)_L \otimes \mathrm{SU}(2)_R } }
\providecommand{\smgroup}{}
\renewcommand{\smgroup}{\ensuremath{\mathrm{SU}(3) \otimes \mathrm{SU}(2)_L \otimes \mathrm{U}(1)_Y } }
\providecommand{\Dpbrane}{}
\renewcommand{\Dpbrane}{\ensuremath{Dp\mathrm{-brane}}}
\providecommand{\Dpbranes}{}
\renewcommand{\Dpbranes}{\ensuremath{Dp\mathrm{-branes}}}
\providecommand{\Dbrane}[1]{}
\renewcommand{\Dbrane}[1]{\ensuremath{D#1\mathrm{-brane}}}
\providecommand{\Dbranes}[1]{}
\renewcommand{\Dbranes}[1]{\ensuremath{D#1\mathrm{-branes}}}
\providecommand{\vev}{}
\renewcommand{\vev}{VEV}
\providecommand{\vevs}{}
\renewcommand{\vevs}{VEVs}
\providecommand{\cm}{}
\renewcommand{\cm}{Constable-Myers }
\providecommand{\Tr}{}
\renewcommand{\Tr}{\ensuremath{\mathrm{Tr}}}
\providecommand{\ads}{}
\renewcommand{\ads}{\ensuremath{\mathrm{AdS_5\times S^5}} }
\providecommand{\adsend}{}
\renewcommand{\adsend}{\ensuremath{\mathrm{AdS_5\times S^5}}}
\providecommand{\identity}{}
\renewcommand{\identity}{\ensuremath{\mathrm{I}}}
\providecommand{\Kahler}{}
\renewcommand{\Kahler}{K\"ahler}
\providecommand{\Schrodinger}{}
\renewcommand{\Schrodinger}{Schr\"odinger\,}
\providecommand{\Fbar}{}
\renewcommand{\Fbar}{\overline{F}}
\providecommand{\GSM}{}
\renewcommand{\GSM}{\ensuremath{G_{\mathrm{SM}}}}
\providecommand{\dotted}[1]{}
\renewcommand{\dotted}[1]{\ensuremath{\stackrel{.}{#1}}}
\providecommand{\Poincare}{}
\renewcommand{\Poincare}{Poincar\'e }
\providecommand{\Angstrom}{}
\renewcommand{\Angstrom}{$\mathring{A}$ngstrom }
\providecommand{\Tr}{}
\renewcommand{\Tr}{\mbox{Tr\,}}
\providecommand{\qq}{}
\renewcommand{\qq}{\ensuremath{\mathrm{\bar{q}q}\,} }
\providecommand{\beq}{}
\renewcommand{\beq}{\begin{equation}}
\providecommand{\eeq}{}
\renewcommand{\eeq}{\end{equation}}
\providecommand{\bea}{}
\renewcommand{\bea}{\begin{eqnarray}}
\providecommand{\eea}{}
\renewcommand{\eea}[1]{\label{#1}\end{eqnarray}}
\providecommand{\Re}{}
\renewcommand{\Re}{\mbox{Re}\,}
\providecommand{\Im}{}
\renewcommand{\Im}{\mbox{Im}\,}
\providecommand{\yms}{}
\renewcommand{\yms}{\ensuremath{YM^*\,}}
\providecommand{\slsh}{}
\renewcommand{\slsh}[1]{\ensuremath{\not\!\!#1}}
\providecommand{\Dsl}{}
\renewcommand{\Dsl}{\ensuremath{\slsh{D}}}
\providecommand{\qb}{}
\renewcommand{\qb}{\ensuremath{\bar{q}{q}} }
\providecommand{\qbend}{}
\renewcommand{\qbend}{\ensuremath{\bar{q}{q}}}
\providecommand{\omp}{}
\renewcommand{\omp}{\ensuremath{0^{-+}\ }}
\providecommand{\opp}{}
\renewcommand{\opp}{\ensuremath{0^{++}\ }}
\providecommand{\nee}{}
\renewcommand{\nee}{${\cal N}=8$ gauged supergravity\ }
\providecommand{\neeend}{}
\renewcommand{\neeend}{${\cal N}=8$ gauged supergravity}
\providecommand{\nefour}{}
\renewcommand{\nefour}{\ensuremath{{\cal N}=4\ }}
\providecommand{\neeight}{}
\renewcommand{\neeight}{\ensuremath{{\cal N}=8\ }}
\providecommand{\netwos}{}
\renewcommand{\netwos}{\ensuremath{{\cal N}=2^\star\ }}

\def\N{{\cal N}}


\thispagestyle{empty}
\renewcommand{\thefootnote}{\fnsymbol{footnote}}

\bigskip

\begin{center} \noindent \Large \bf
Canonical Coordinates and Meson Spectra for Scalar Deformed ${\cal
N}=4$ SYM from the AdS/CFT Correspondence
\end{center}

\bigskip\bigskip\bigskip

\centerline{ \normalsize \bf Jonathan P. Shock
\footnote[1]{\noindent \tt
 jps@itp.ac.cn} }

\bigskip
\bigskip\bigskip

\centerline{ \it Institute of Theoretical Physics} \centerline{
\it  Chinese Academy of Sciences} \centerline{\it   P.O. Box 2735
} \centerline{ \it  Beijing 100080, CHINA}
\bigskip

\bigskip\bigskip

\renewcommand{\thefootnote}{\arabic{footnote}}

\centerline{\bf \small Abstract}
\medskip

{\small \noindent Two points on the Coulomb branch of ${\cal N}=4$
super Yang Mills are investigated using their supergravity duals.
By switching on condensates for the scalars in the \nefour
multiplet with a form which preserves a subgroup of the original
$R$-symmetry, disk and sphere configurations of D3-branes are
formed in the dual supergravity background. The analytic,
canonical metric for these geometries is formulated and the
singularity structure is studied. Quarks are introduced into the
corresponding field theories using D7-brane probes and the meson
spectrum is calculated. For one of the condensate configurations,
a mass gap is found and shown analytically to be present in the
massless limit. It is also found that there is a stepped spectrum
with eigenstate degeneracy in the limit of small quark masses and
this result is shown analytically. In the second, similar
deformation it is necessary to understand the full D3-D7 brane
interaction to study the limit of small quark masses. For quark
masses larger than the condensate scale the spectrum is calculated
and shown to be discrete as expected.

\newpage


\section{Introduction}

Substantial effort has gone into understanding the properties of
gauge theories dual to supergravity backgrounds which asymptote to
\ads. This progress has been made possible by the conjectured
AdS/CFT correspondence
\cite{hep-th/9711200,hep-th/9802109,hep-th/9802150}. In
particular, there has been much interest in attempting to obtain
QCD-like models by breaking the supersymmetry and conformal
symmetry via the inclusion of relevant operators
\cite{hep-th/9903026,hep-th/9909047,hep-th/0004063,hep-th/0003136}.
The addition of quarks \cite{hep-th/0205236, hep-th/0211107,
hep-th/0304032, hep-th/0305049, hep-th/0306018, hep-th/0311201,
hep-th/0311084, hep-th/0307218, hep-th/0312071, hep-th/0403279,
hep-th/0406207, hep-th/0408113, hep-th/0410035, hep-th/0411097,
hep-th/0412260, hep-th/0412141,
hep-th/0502091,hep-th/0511044,hep-th/0511045} has also heralded a
great leap in creating realistic models and has allowed us to
calculate many non-perturbative quantities. Over the last year
several toy models of five-dimensional holography have also made
progress in describing theories with a small number of colours and
even the most naive of these scenarios appears to give remarkable
agreement with lattice QCD and experimental observations of meson
masses and decay constants
\cite{hep-ph/0501128,hep-ph/0501218,hep-th/0501022,hep-ph/0510334,hep-ph/0603142,hep-ph/0603249}.
Generally, even in the simplest deformations, calculations must be
performed numerically in order to find solutions to five and
ten-dimensional equations of motion. Similar models of AdS slices
were also considered in \cite{hep-th/0209080,hep-th/0212207} where
glueball spectra were studied.

In this paper we study one particular deformation \cite{hep-th/9906194} of the \ads geometry using D7-brane probes. The field theory dual to this geometry retains the full ${\cal N}=4$
supersymmetry but breaks the $SU(4)_R$ symmetry by the addition of
condensates for the three complex scalar fields. Two point
correlation functions and Wilson loops have been studied in this theory on the Coloumb branch of ${\cal N}=4$ Super
Yang-Mills and the features of the scalar spectrum and
screening have been explained in terms of ensembles of brane
distributions. In the current work, the supersymmetry allows the analytic
form of the metric to be obtained which encodes the field theory
in its canonically normalised form, a task which is more difficult in non-supersymmetric backgrounds. In the canonically normalised coordinates the meson spectrum can be calculated as a function of quark mass once D7-branes have been introduced.

The addition of fundamental matter to the theory under consideration has been studied briefly in \cite{hep-th/0502091}. When the D7-brane flow was calculated in this geometry, there appeared to be a small but finite quark bilinear condensate for non-zero quark mass. This result is surprising because this supersymmetric theory should not support a chiral condensate. This
anomalous result is however due to the use of an unsuitable basis
with which to describe the geometry in a holographic setting. The details of this are addressed in section \ref{sec.cancon}. It
is trivial to prove that any supersymmetric background when
written in canonical form will define a stable field theory with
zero vacuum expectation value for the quark bilinears. It was
shown using this background as a simple example that the original,
geometric interpretation of chiral symmetry breaking was not
sufficient for the analysis of backgrounds out of their canonical
form. A method was developed by which the potential felt by a
D7-brane in the singular region could be studied out of canonical
form. It is the aim of the current paper to find the canonical basis with which to describe the field theory and study
the spectrum therein.

\section{The Supergravity Geometries}

The two geometries of interest in this paper are two of a set of
five solutions discussed in \cite{hep-th/9906194} which are all
asymptotically \ads and are sourced purely by D3-branes. Each
background is formulated in terms of a D3-brane density
distribution function.

It is possible to find the analytic, canonical form for all five
of these supergravity backgrounds though only in two of the cases
is the form of the metric simple enough to calculate the meson
spectrum.

Each background is dual to an ${\cal N}=4$ field theory with a scalar
condensate, preserving a subgroup of the original $SU(4)_R$
symmetry. In each case the metric is given by
\begin{equation}
ds^2=\frac{1}{\sqrt{H}}dx_{\cal M}^2-\sqrt{H}\sum_{i=1}^6 dy_i^2\
,
\end{equation}
where the warp factor is
\begin{equation}
H=\int_{|\vec{\omega}|<l}d^n\omega\sigma(\vec{\omega})\frac{L^4}{|\vec{y}-\vec{\omega}|^4}\
.
\end{equation}
$\vec{\omega}$ is a vector in $n$ of the six dimensions transverse
to the D3-brane worldvolume and $\vec{y}$ is a vector in all six
of these dimensions. $l$ parametrises the size of the D3-brane
distribution and is the single, extra, free parameter in each of
these geometries. The integral is performed over the region of
space with support from the distribution function. The dimension
of the distribution, $n$, together with the density function, the
preserved symmetry and the form of the scalar condensate are
provided in table \ref{tab.densdist}.
\begin{table}[!h]
\begin{center}
\begin{tabular}{|c|c|c|c|}
  \hline
  $n$ & $\sigma(\vec{\omega})$ & Preserved Symmetry & Scalar Condensate \\
  \hline
  \hline
  1 & $\frac{2}{\pi l^2}\sqrt{l^2-\omega^2}$ & $SO(5)$ & $\frac{1}{15}(1,1,1,1,1,-5)$ \\
  2 & $\frac{1}{\pi l^2}\theta(l^2-\omega^2)$ & $SO(4)\times SO(2)$ & $\frac{1}{6}(1,1,1,1,-2,-2)$ \\
  3 & $\frac{1}{\pi^2 l^2}\frac{1}{\sqrt{l^2-\omega^2}}$ & $SO(3)\times SO(3)$ & $\frac{1}{3}(1,1,1,-1,-1,-1)$ \\
  4 & $\frac{1}{\pi^2 l^2}\delta(l^2-\omega^2)$ & $SO(2)\times SO(4)$ & $\frac{1}{6}(2,2,-1,-1,-,1-,1)$ \\
  5 & $\frac{1}{\pi^3 l^2}\left(\frac{\delta(l^2-\omega^2)}{\sqrt{l^2-\omega^2}}-\frac{\theta(l^2-\omega^2)}{2(l^2-\omega^2)^\frac{3}{2}}\right)$ & $SO(5)$ & $\frac{1}{15}(5,-1,-1,-1,-1,-1)$ \\
  \hline
\end{tabular}
\end{center}
\caption{D3-brane density distribution functions preserving
subgroups of the $SO(6)$ symmetry. The condensate is a vector in
the six-dimensional space of real scalars of ${\cal N}=4$
SYM.}\label{tab.densdist}
\end{table}

\section{Obtaining the Canonical Coordinates}\label{sec.can}

The two backgrounds of interest are those which preserve $SO(2)\times SO(4)$ subgroups of the original $SO(6)$ symmetry. These are particularly interesting because once fundamental matter is added, the chiral symmetry is described explicitely by the $SO(2)$ geometrical symmetry.

A non-canonical analytic form for these backgrounds has been given
in \cite{hep-th/9906194}. From this form of the metric it is simple
to find the canonical system. A D3-brane is introduced which
describes a field theory with six scalar fields. The canonical
metric is defined as that in which the six scalar fields are
simultaneously canonically normalised.\footnote{Thanks to K Sfetsos
for pointing out that this canonical coordinate system has been
calculated previously in \cite{hep-th/9811167} and
\cite{hep-th/9906201}}.}

Note that recent work \cite{hep-th/0512125} describes an
unambiguous method of finding the natural coordinate system for
supersymmetric deformations using holographic renormalisation.

One of the backgrounds of interest dual to ${\cal N}=4$ SYM with
an adjoint scalar condensate has been studied previously \cite{hep-th/0105235, hep-th/0502091} in the
context of flavoured holographic models. This geometry was
conjectured to be equivalent to the $n=2$ geometry of table
\ref{tab.densdist}, though we show in this section that it is
really another parametrisation of the $n=4$ deformation.

The original coordinate system used to study mesons in this
background is a limiting case of the ${\cal N}=2$ background of
\cite{hep-th/0004063}. To return to the full ${\cal N}=4$ theory,
the two five-dimensional supergravity scalar fields are equated to
acquire a theory with six scalar vevs of the form in table
\ref{tab.densdist}. In this particular, limiting case of the supergravity solution, the dilaton becomes constant. It may be interesting
to study those geometries with a running dilaton in their canonical coordinates in the future.

In this supergravity background, there are two fields of interest.
One is the five-dimensional scalar field, $\chi$, and the other is
the warp factor, $A$, multiplying the Minkowski space-time
components in the five-dimensional truncation of the
ten-dimensional metric. In \cite{hep-th/9906194} the lift of the
five-dimensional supergravity theory was obtained and we use the
resulting metric in what follows.

The equations of motion for the scalar field and warp factor are
\begin{equation}\label{eq.foeom}
\frac{d\chi}{du}=\frac{1}{3R}\left(\frac{1}{\chi}-\chi^5\right)\ ,
\end{equation}
and
\begin{equation}\label{eq.Achieq}
e^{2A}=\frac{l^2}{R^2}\frac{\chi^4}{\chi^6-1}\ .
\end{equation}

\begin{figure}[!h]
\begin{center}
\includegraphics[width=10cm,clip=true,keepaspectratio=true]{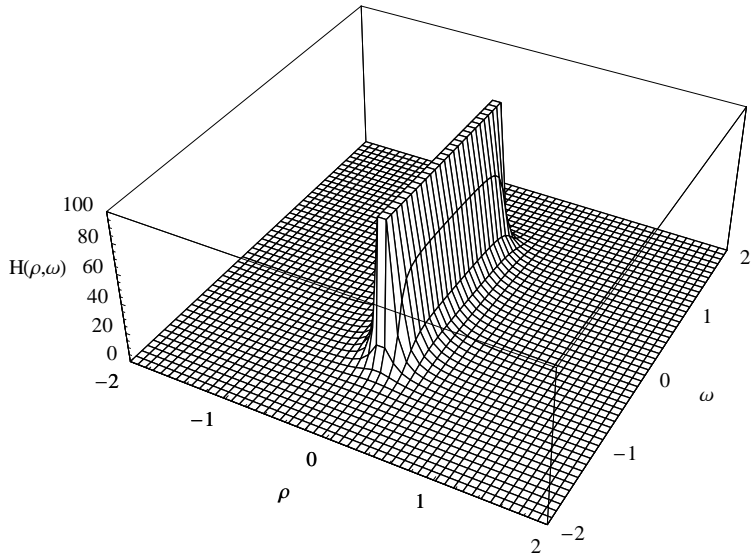}
\caption{Singularity structure of the warp factor for the $n=4$
geometry in the $(\omega,\rho)$ plane with
$l^2=-1$.}\label{fig.harm1}
\end{center}
\end{figure}

The metric for this background in the unphysical coordinates is
given by
\begin{equation}\label{eq.unphys}
ds^2=\frac{\sqrt{X}}{\chi}e^{2A}dx_{\cal
M}^2-\frac{\sqrt{X}}{\chi}\left(du^2+\frac{R^2}{\chi^2}\left(d\theta^2+\frac{\sin^2\theta}{X}d\phi^2+\frac{\chi^6\cos^2\theta}{X}d\Omega_3^2\right)\right)\
,
\end{equation}
where
\begin{equation}
X=\cos^2\theta+\chi^6\sin^2\theta\ .
\end{equation}
By probing with a D3-brane, the action for the six scalar fields
is seen not to be canonically normalised, though the moduli space
is manifest. In this case it is possible to use the first order
supergravity equations of motion to find the correct coordinate
system in which to describe the field theory in its canonical
form. The result of this transformation is given by
\begin{equation}\label{eq.neq2}
ds^2=H^{-\frac{1}{2}}dx_{\cal
M}^2-H^\frac{1}{2}\left(d\rho^2+ d\omega_5^2+d\omega_6^2+\rho^2 d\Omega_3^2\right)\ ,
\end{equation}
where the warp factor is
\begin{equation}
H(\rho,\omega,l)=\frac{2R^4}{l^4+2l^2(\omega^2-\rho^2)+(\omega^2+\rho^2)^2+(l^2+\omega^2+\rho^2)\sqrt{(\omega^2+(l-\rho)^2)(\omega^2+(l+\rho)^2)}}\
.
\end{equation}
H is plotted in the $(\omega,\rho)$ plane in figure
\ref{fig.harm1} in order to illuminate the singularity structure.
This deformation corresponds to an $S^3$ distribution of D3-branes
spanning the locus $\rho=l$ in the $\mathbf{R}^4=S^3\times \rho$
plane. This is therefore the analytic, canonical form of the $n=4$
metric in \cite{hep-th/9906194} which describes an ${\cal N}=4$
field theory with vacuum expectation values for all six scalar
fields in the configuration $(2,2,-1,-1,-1,-1)$. The original
metric also encodes the $n=2$ solution with scalar vev
$(1,1,1,1,-2,-2)$, however this configuration has negative $l^2$
as discussed in \cite{hep-th/9906194}.

\begin{figure}[!h]
\begin{center}
\includegraphics[width=10cm,clip=true,keepaspectratio=true]{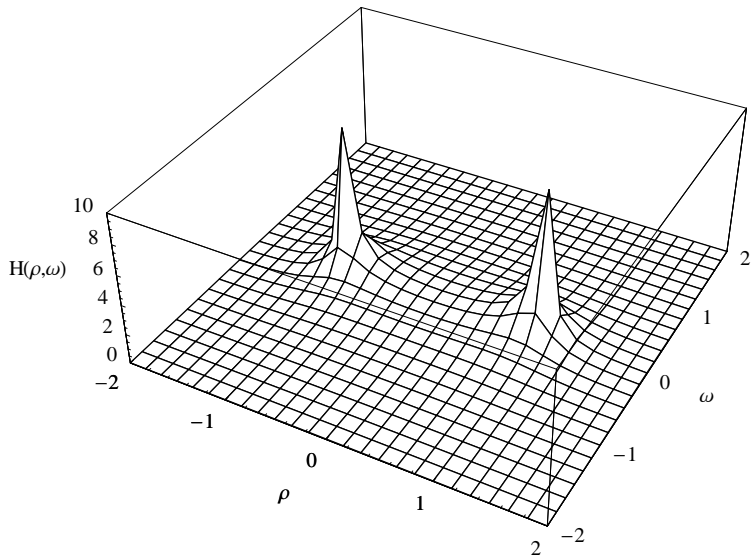}
\caption{Singularity structure of the warp factor for the $n=2$
geometry in the $(\omega,\rho)$ plane as in figure \ref{fig.harm1}
but with $l^2=1$.}\label{fig.harm2}
\end{center}
\end{figure}

The warp factor for the $n=2$ configuration can be plotted in the
same way and is illustrated in figure \ref{fig.harm2}.
This corresponds to a $D^2$ distribution spanning the $(\omega_5,
\omega_6)$ two-plane.
It is possible to calculate the canonical, analytic form for the
other three metrics in \cite{hep-th/9906194} which preserve
different subgroups of the original $SO(6)$ symmetry. However, the
analytic form for these backgrounds are too complicated to calculate
the meson spectra so will be discussed no further.

\section{Mesons from D7-Brane Probes}
Having found the correct coordinate system in which to describe a
canonically normalised field theory living on the D3-branes, we
can study the theory with the addition of quarks.

We start with the background described by equation \ref{eq.neq2} with positive $l^2$,
which preserves an $SO(4)\times SO(2)$ subgroup of the original
symmetry. The D7-brane is embedded by filling the four Minkowski
space-time directions, the $\rho$-direction and wrapping the
$S^3$. This wrapped cycle ensures the stability of the brane
configuration.

Before calculating the $x$-dependent excitations of the brane, we
must study its flow purely in the $\rho$-direction. This
corresponds to calculating $\omega_5$ and $\omega_6$ as a function
of $\rho$. However, because of the supersymmetric nature of this
background, the warp factor in front of the $\mathbf{R}^6$ is the
inverse of that in front of the ${\cal M}^4$ plane. This means
that the $x$-independent flow is exactly the same as in the \ads
background to which the solution is known analytically.

We can solve the equation of motion and obtain the following
solutions.
\begin{equation}
\omega=\int \frac{c_1}{\sqrt{\rho^6-c_1^2}}d\rho+c_2\ ,
\end{equation}
where $\omega=\sqrt{\omega_5^2+\omega_6^2}$. It is clear that this
function will not be real all the way to $\rho=0$ for $c_1\ne 0$.
The physical solution, corresponding to the renormalisation group
flow of the brane, must therefore be $\omega=c_2$, equivalent to a
quark mass but no quark bilinear condensate. This will always be
the case for a supersymmetric solution where the warp factors
cancel leaving the \ads equation of motion (excluding
$x$-dependent fluctuations).

Because of the manifest $SO(2)$ symmetry between $\omega_5$ and
$\omega_6$, we are free to choose the direction in which to
explicitly break this rotational invariance. For simplicity, we
choose the solution $\omega_5=m$ and $\omega_6=0$. We now want to
study the mesonic fluctuations about this brane flow. We study the
modes in the $\omega_6$-direction given by
$\omega_6=0+\tilde{\omega}_6(\rho,x)$. Note that because there is
no chiral symmetry breaking in this background, the mesonic
excitations in the $\omega_5$ and $\omega_6$-directions will be
identical. This means that the positive and negative parity states
will be degenerate.

The action for $\tilde{\omega}_6$ up to quadratic order
is given by
\begin{eqnarray}
S&=&\int d^8\zeta \rho^3\left(\left({\frac{\partial
\tilde{\omega}_6}{\partial
\rho}}\right)^2\right. \nonumber\\
&+&\left. \left({\frac{\partial \tilde{\omega}_6}{\partial
x}}\right)^2\frac{2R^4}{l^4+2l^2(m^2-\rho^2)+(m^2+\rho^2)^2+(l^2+m^2+\rho^2)\sqrt{(m^2+(l-\rho)^2)(m^2+(l+\rho)^2)}}\right)\ .\nonumber\\
\end{eqnarray}
For small oscillations about the flow $\omega_5(\rho)$, the meson
interaction terms will be subdominant and the function
$\tilde{\omega}_6$ can be treated as a plane wave in the Minkowski
space-time directions and therefore the ansatz for this function
is given by
\begin{equation}
\tilde{\omega}_6=f(\rho)e^{ik.x}\ .
\end{equation}
This ansatz which is independent of the three-sphere coordinates
corresponds to an $R$-singlet, spinless meson wavefunction. The
equation of motion for $f(\rho)$ is given by
\begin{equation}\label{eq.eomf}
\frac{2M^2 \rho
R^4f(\rho)}{l^4+2l^2(m^2-\rho^2)+(m^2+\rho^2)^2+(l^2+m^2+\rho^2)\sqrt{(m^2+(l-\rho)^2)(m^2+(l+\rho)^2)}}+3f'(\rho)+\rho
f''(\rho)=0\ ,
\end{equation}
where $M^2=-k^2$. The eigenvalues, $M$, are given by the values
for which the flow of $f$ is well behaved all the way to $\rho=0$
and normalisable in the UV. It must also have the correct scaling
dimensions in the \ads limit to describe a mesonic excitation. Of
the two UV solutions, the solution corresponding to meson
fluctuations, as opposed to an $x$-dependent mass is
$f(\rho)\rightarrow \frac{c}{\rho^2}$.

It is possible to remove the explicit $R$ and $l$ dependence in the above equation of motion by performing the following rescaling:

\begin{equation}
m=\tilde{m}l,\hspace{0.5cm}\rho=\tilde{\rho}l,\hspace{0.5cm}MR^2=\tilde{M}l,
\end{equation}
giving us the following equation of motion
\begin{equation}\label{eq.eomf2}
\frac{2\tilde{M}^2 \tilde{\rho}
f(\tilde{\rho})}{1+2(\tilde{m}^2-\tilde{\rho}^2)+(\tilde{m}^2+\tilde{\rho}^2)^2+(1+\tilde{m}^2+\tilde{\rho}^2)\sqrt{(\tilde{m}^2+(1-\tilde{\rho})^2)(\tilde{m}^2+(1+\tilde{\rho})^2)}}+3f'(\tilde{\rho})+\tilde{\rho}
f''(\tilde{\rho})=0\ ,
\end{equation}

\begin{figure}[!h]
\begin{center}
\includegraphics[width=12cm,clip=true,keepaspectratio=true]{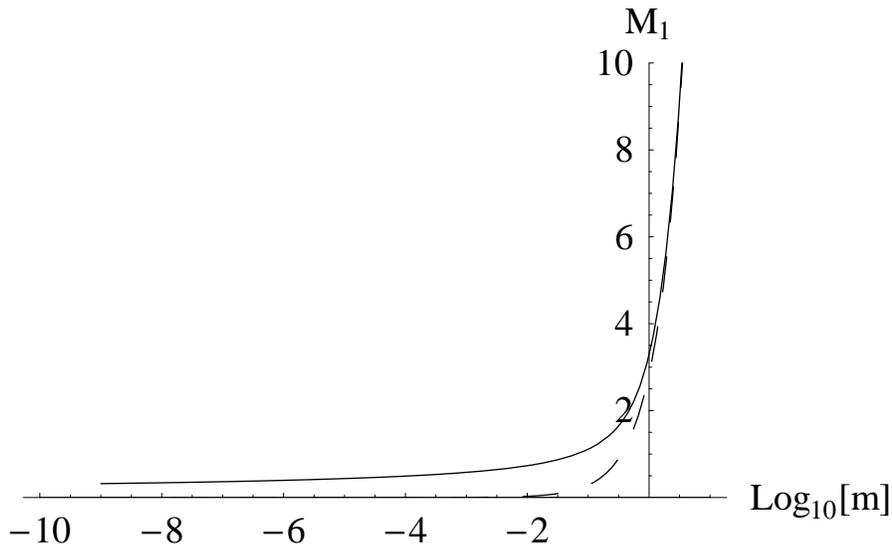}
\caption{$M_1$ against $\log_{10}m$ for $l=1,R=1$ (solid line). This
becomes exactly degenerate with $M_1$ for $l=0,R=1$ (dashed line) in
the large $m$ limit. Note the appearance of a mass gap. The scale is
set by the AdS radius $R$ which can be tuned by hand to compare with
lattice data.}\label{fig.logmM}
\end{center}
\end{figure}

Before performing this rescaling we can take the $l\rightarrow 0$ limit and find that the numerical values coincide with
the known analytic results of the pure \ads spectrum
\cite{hep-th/0311270}:
\begin{equation}\label{eq.adsspectr}
M=\frac{2m}{R^2}\sqrt{(n+1)(n+2)}\ , \hspace{1cm} n=0,1,...
\end{equation}
showing that in this limit the numerics are under control. For the
rescaled equation (eq.\ref{eq.eomf2}) we can study the lowest mass
meson as a function of the quark mass. Figure \ref{fig.logmM}
shows the value of the first meson mass as a function of the quark
mass. The important point to note here is that there appears to be
a mass gap in the $m\rightarrow 0$ limit for $l\ne 0$. The
numerics make this calculation difficult, though at $m=10^{-10}$
the value of $M_1$ is $0.28$. Note that in contrast to the
equation of motion with $l=0$, the D7-brane equation is perfectly
well behaved in this limit and has discrete eigenvalues. This will
be shown analytically in section \ref{sec.an}.

\begin{figure}[!h]
\begin{center}
\includegraphics[width=12cm,clip=true,keepaspectratio=true]{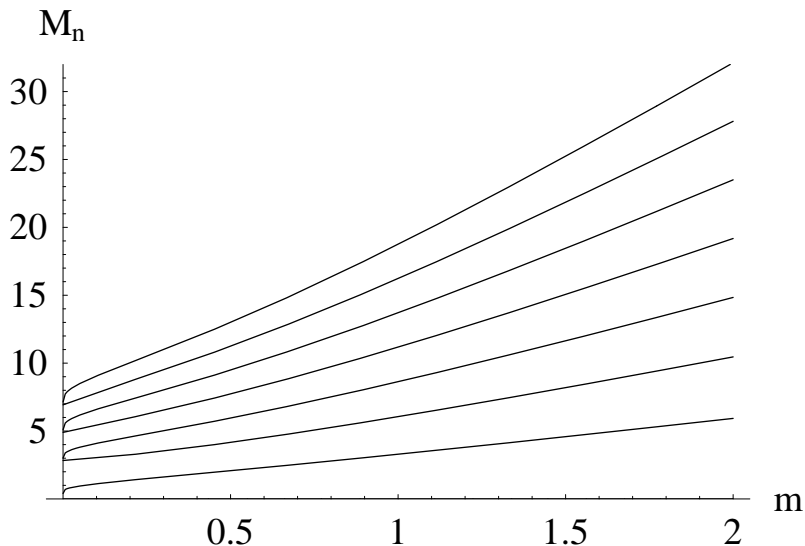}
\caption{$M_n$ against $m$ for the first seven states. For large
quark masses the behaviour is simple as the brane cannot resolve the
double singularity structure, in the small mass limit there is a
degeneracy of states. In the large quark mass limit the spectrum
goes like $M R^2 =2m\sqrt{(n+1)(n+2)}$ just as in the pure \ads
case.}\label{fig.spectra}
\end{center}
\end{figure}

Having calculated the first meson mass, we can study the spectrum
for the higher excited states. The results of this are shown in
figure \ref{fig.spectra}.
We can see from figure \ref{fig.spectra} that there is a
degeneracy for quark masses much less than the condensate scale
(given by the gap between the singularities in the supergravity
geometry). For large quark masses, corresponding to the D7-brane
lying far from the singular region, there is a spectrum which is
given by
\begin{equation}
M R^2 =2m\sqrt{(n+1)(n+2)}, \hspace{0.5 cm} n=0,1,...
\end{equation}
as expected because the brane cannot resolve the separated
singularities for large quark masses. For smaller masses however we
see that pairs of excited states become degenerate and indeed there
appears to be an exact degeneracy as we get to the massless limit.
The lowest state however does not appear to have a degeneracy and we
will show in section \ref{sec.an} that this mode does not appear in
the exact massless limit when we study the equations analytically.

\subsection{Significance of the Canonical Coordinates}\label{sec.cancon}

It is important at this stage to note why we are interested in
finding one particular coordinate system for this problem. Though it
appears strange that coordinates matter in what is essentially an
eigenvalue problem we see in the following that this canonical
coordinate system simplifies the calculations considerably. Without
the change of coordinates not only would we be unable to obtain an
analytic expression for the meson spectrum in the massless quark
limit but because of the numerical behaviour of the supergravity
field, $\chi$, we would not be able to obtain a reliable spectrum at
all. This means that the numerical instabilities are critical
to the eigenvalue problem.

These instabilities stem from two regions in the geometry. The first
of these is the singular region where the supergravity field,
$\chi$, asymptotes to infinity. The second region is where the space
returns to pure \adsend. In this region the supergravity field asymptotes
to unity but the further we go into the ultraviolet, the more
significant becomes the accuracy of the solution in insuring a
numerically stable geometry. This necessity for ever higher
numerical accuracy in the ultraviolet can be seen in equations
\ref{eq.foeom} and \ref{eq.Achieq}. In the canonical basis the
singular behaviour cancels explicitly in the flow equation for the
D7-brane.

\begin{figure}[!h]
\begin{center}
\includegraphics[width=12cm,clip=true,keepaspectratio=true]{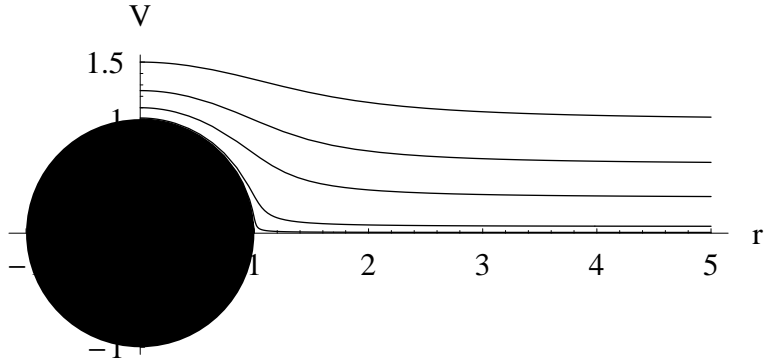}
\caption{Stable D7-brane flows for varying quark masses. For ease of
visualisation the flows are plotted using the coordinates
$r^2+v^2=R^2e^{2u}$ and
$\frac{v}{r}=\tan{\theta}$ with a singularity at radius $R$.}\label{fig.neq4flows}
\end{center}
\end{figure}

\begin{figure}[!h]
\begin{center}
\includegraphics[width=10cm,clip=true,keepaspectratio=true]{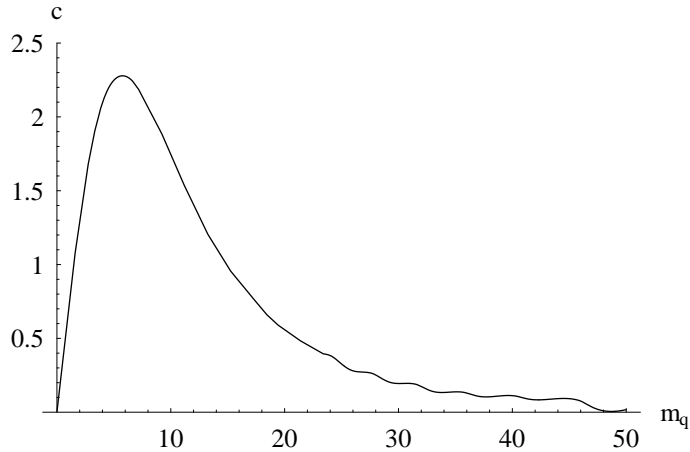}
\caption{Quark bilinear condensate versus quark mass calculated from
the stable D7-brane flows in figure
\ref{fig.neq4flows}.}\label{fig.Cvsm}
\end{center}
\end{figure}

The first hint that these instabilities are critical is seen when we
calculate the stable value of the quark bilinear condensate as a
function of the quark mass. The stable flows of the D7-brane for
various quark masses and the quark bilinear condensate versus quark
mass are plotted in figures \ref{fig.neq4flows} and \ref{fig.Cvsm}
respectively.
In the canonical coordinates we see analytically that there is no
condensate present for any quark mass. However, in the original
coordinate system (equation \ref{eq.unphys}), we see that for finite $r$ the stable D7-brane
solution does not lie flat with respect to the $r$ axis indicating a condensate present for non-zero quark mass. In theory we should be calculating the
condensate at infinite energy, where the space returns to pure \ads.
This is not possible as we have had to solve the equation
numerically. Solving for finite $r$ will always give a non-zero
condensate for non-zero quark mass. As we perform the calculation
further into the UV where this calculation should be more accurate,
the numerical instabilities of the scalar supergravity field become
critical and a meaningful result becomes harder to calculate.

When calculating the meson spectrum as a function of the quark mass,
the spectrum which is obtained is also found to be a chaotic one,
clearly influenced by the singular behaviour in the IR and the
numerical instabilities in the UV, especially in the region of small
quark mass which is the domain of most interested.
This instability is critical to the calculation and means that the
interesting properties which can be found analytically and are
discussed in the next section are not observable in the original basis using the computational techniques employed here.

\subsection{Analytical Results}\label{sec.an}
Though we know from the $m\rightarrow 0$ limit of equation
\ref{eq.adsspectr} (The \ads meson spectrum) that at $m=0$ the spectrum becomes continuous,
this appears not to be the case from the numerical calculation in
the $l\ne 0$ deformation. In this case, we can take the $m=0$
limit explicitly in the equation of motion and retain a discrete
spectrum.
\begin{equation}
M^2 \rho R^4 f(\rho)+l^2(l^2-\rho^2)(3f'(\rho)+\rho f''(\rho))=0\
.
\end{equation}
The solution to this equation is given in
\cite{hep-th/9906194,hep-th/9906201} for the case of the glueball
spectrum and is given by:
\begin{equation}
MR^2=2l\sqrt{(n+1)(n+2)}\, \hspace{1cm} n=0,1,...
\end{equation}
(see also \cite{hep-th/0510132} for higher angular momentum
eigenvalues). It is also precisely the same spectrum as the \ads
spectrum with an interchange of the quark mass, $m$ and the
deformation parameter, $l$ (the magnitude of the scalar condensate).
It is particularly interesting that at this point on the moduli
space the spectrum of ${\cal N}=2$ quarks has the same spectrum of
states as the ${\cal N}=4$ theory.

There is an apparent discrepancy between the analytical and
numerical values calculated in the $m\rightarrow 0$ limit.
Interestingly, the masses using the numerical and analytic methods
are exactly equal except for the very first state. However,
because the infinitesimal and exact massless limits are
qualitatively different we may expect a discrete change in
behaviour.

For the numerical calculation, in the case,
$m=10^{-10}$ and $l=R=1$, the spectrum is given by the values in table \ref{tab.numerical}. The degeneracy is given by the number of mesons with
approximately the same eigenvalue.
\begin{table}[!h]
\begin{center}
\begin{tabular}{||c|c||}
  \hline
  M & Degeneracy \\
  \hline
  0.28 & 1 \\
  2.8 & 2 \\
  4.9 & 2 \\
  6.9 & 2 \\
  8.9 & 2 \\
  \hline
\end{tabular}
\caption{The mass spectrum for $l=R=1$ and $m\rightarrow 0$
calculated numerically. The degeneracy is given by the number of
states with converging eigenvalues.}\label{tab.numerical}
\end{center}
\end{table}

This should be compared with the exact $m=0$ result obtained
analytically where the degeneracy is not obtainable. These results are given in table \ref{tab.analytical}.
\begin{table}[!h]
\begin{center}
\begin{tabular}{||c|c||}
  \hline
  M & Degeneracy \\
  \hline
  2.8 & n.a \\
  4.9 & n.a \\
  6.9 & n.a \\
  8.9 & n.a \\
  \hline
\end{tabular}
\caption{The mass spectrum for $l=R=1$ and $m=0$ calculated
analytically. The degeneracy is not possible to study in the
analytic massless limit because it is not possible to calculate
the number of eigenstates of a given
eigenvalue.}\label{tab.analytical}
\end{center}
\end{table}

\subsection{The $n=2$ Spectrum}
\begin{figure}[!h]
\begin{center}
\includegraphics[width=12cm,clip=true,keepaspectratio=true]{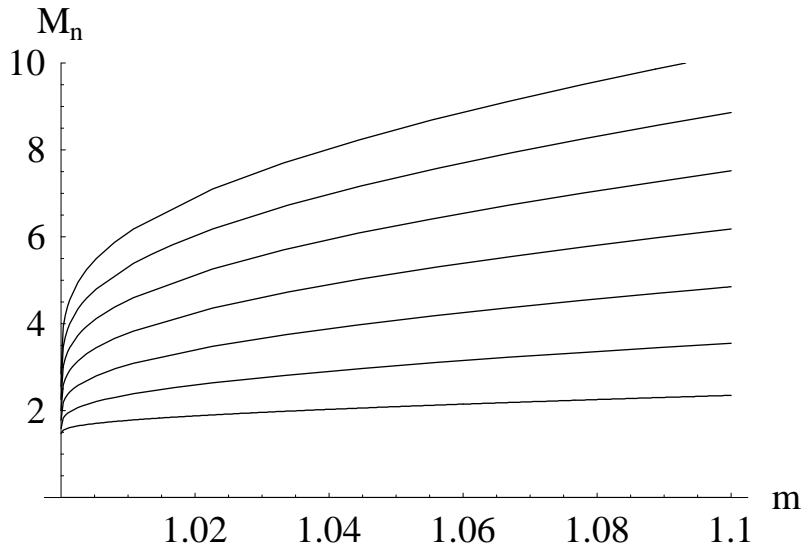}
\caption{$M_n$ against $m$ for the first seven states of the $n=2$
deformation for $m>l$ (solid lines).}\label{fig.spectran2}
\end{center}
\end{figure}

As discussed in section \ref{sec.can} this deformation is obtained
from the $n=4$ deformation with the interchange
$l^2\rightarrow-l^2$. This transformation however has a
significant effect on the nature of this geometry from the point
of view of a D7-brane probe. Again, we can see from the simple
product form of the geometry with inverse warp factors for the
${\cal M}^4$ and $\mathbf{R}^6$ that to zeroth order in mesonic
excitations, the D7-branes will lie flat and not notice the
singular structure. In figure \ref{fig.harm2}, it was shown that
there is a disk singularity lying in the $(\omega_5,\omega_6)$
plane of radius $l$. This means that a brane corresponding to
adding quarks of mass $m$ for $m<l$ will pass straight through the
singular region. We know that this is not a physical configuration
and the full interaction between the D3-brane stack and the
D7-brane probe would be needed to understand this case fully. It
is however possible to study quarks with $m>l$ in the
supergravity, probe approximation. We find that as expected, for
large masses, the spectrum returns to the \ads values, just as in
the large mass limit of the $n=4$ geometry. The spectrum down to
$m=l$ is provided in figure \ref{fig.spectran2}
Though it appears from this diagram that the states may become
degenerate (and possibly massless) at $m=l$ this is not apparently
the case from these numerical computations, The first seven states
are given by the following values at $m=l$:
\begin{equation}
M_n=\{1.46, 1.59, 1.77, 2.00, 2.26, 2.54, 2.84\}
\end{equation}
In fact the equations are perfectly well behaved for $m<l$ and the
spectrum is calculable in this region, however the results should
not be trusted as clearly interactions between the D7-brane and
the singular D3-brane distribution must be understood fully away
from the supergravity limit.

We find that $M_n$ is linear in $l$ for $m=l$, though the
$n$-dependence is not as simple as the massless $n=4$ geometry.
Although the equation of motion simplifies significantly to
\begin{equation}
\frac{2\tilde{M}^2f[\tilde{\rho}]}{4+\tilde{\rho}^2+\tilde{\rho}\sqrt{4+\tilde{\rho}^2}}+3\tilde{\rho}f'[\tilde{\rho}]+\tilde{\rho}^2f''[\tilde{\rho}],
\end{equation}
no analytic solution to this equation was obtained.

The qualitative difference between the meson spectrum in this
geometry and the $n=4$ case is that there does not appear to be a
degenerate spectrum here. The qualitative difference from the point
of view of the D7-brane is that there is a continuous distribution of
branes rather than a set of singular points with a separation of $2l$.

\section{Conclusions}
We have found an analytic, canonical form for two scalar
deformations formulated in \cite{hep-th/9906194}. The singularity
structures are exactly as expected from the D3-brane density
distribution functions. In these two cases the $SO(6)$ symmetry is
broken to $SO(4)\times SO(2)$ which encodes the chiral symmetry of
the field theory explicitly, it is possible to calculate the meson
spectrum from excitations of probe D7-branes. Interestingly, in one
of these backgrounds, it is possible to find the spectrum
analytically in the limit of exactly massless quarks. This result
includes the existence of a mass gap proportional to the deformation
parameter with exactly the same spectrum as the adjoint scalar two
point Greens function of the ${\cal N}=4$ theory of
\cite{hep-th/9906194}. A degeneracy is discovered in the limit of
small quark masses (c.f the deformation parameter $l$).

In the second $SO(4)\times SO(2)$ preserving background, it is
possible to find the spectra for quarks with larger masses, $m$,
than the deformation parameter, $l$. The spectrum in this case
does not appear to be degenerate though to fully understand the
$m\sim l$ limit, higher order corrections to the supergravity
limit must be calculated. The overall conclusion is that by
finding the correct analytic, physical coordinates in which to
describe the dual gravity theory, we can study the elaborate
structure of meson spectra in field theories with complicated
condensate terms switched on. If we can do the same thing in the
non-supersymmetric analogues of these geometries, we may be able
to gain some more insight into real world hadron spectra.

\section{Acknowledgements}

I would like to thank K Sfetsos for pointing out previous, similar
work plus errors in the original version of this paper. I would also
like to thank Nick Evans and Feng Wu for helpful discussions and
clarification. This project is funded by the Project of Knowledge
Innovation Program (PKIP) of the Chinese Academy of Science (CAS).


\begin{thebibliography}{ll}
\bibitem{hep-th/9711200}
  J.~M.~Maldacena,
  Adv.\ Theor.\ Math.\ Phys.\ {\bf 2}, 231 (1998)
  [Int.\ J.\ Theor.\ Phys.\ {\bf 38}, 1113 (1999)]
  [arXiv:hep-th/9711200].

\bibitem{hep-th/9802109}
  S.~S.~Gubser, I.~R.~Klebanov and A.~M.~Polyakov,
  Phys.\ Lett.\ B {\bf 428}, 105 (1998)
  [arXiv:hep-th/9802109].

\bibitem{hep-th/9802150}
  E.~Witten,
  Adv.\ Theor.\ Math.\ Phys.\ {\bf 2}, 253 (1998)
  [arXiv:hep-th/9802150].

\bibitem{hep-th/9903026}
  L.~Girardello, M.~Petrini, M.~Porrati and A.~Zaffaroni,
  JHEP {\bf 9905}, 026 (1999)
  [arXiv:hep-th/9903026].

\bibitem{hep-th/9909047}
  L.~Girardello, M.~Petrini, M.~Porrati and A.~Zaffaroni,
  Nucl.\ Phys.\ B {\bf 569}, 451 (2000)
  [arXiv:hep-th/9909047].

\bibitem{hep-th/0004063}
  K.~Pilch and N.~P.~Warner,
  Nucl.\ Phys.\ B {\bf 594}, 209 (2001)
  [arXiv:hep-th/0004063].

\bibitem{hep-th/0003136}
  J.~Polchinski and M.~J.~Strassler,
  arXiv:hep-th/0003136.

\bibitem{hep-th/0205236}
  A.~Karch and E.~Katz,
  JHEP {\bf 0206}, 043 (2002)
  [arXiv:hep-th/0205236].

\bibitem{hep-th/0211107}
  A.~Karch, E.~Katz and N.~Weiner,
  Phys.\ Rev.\ Lett.\ {\bf 90}, 091601 (2003)
  [arXiv:hep-th/0211107].

\bibitem{hep-th/0304032}
  M.~Kruczenski, D.~Mateos, R.~C.~Myers and D.~J.~Winters,
  JHEP {\bf 0307}, 049 (2003)
  [arXiv:hep-th/0304032].

\bibitem{hep-th/0305049}
  T.~Sakai and J.~Sonnenschein,
  JHEP {\bf 0309}, 047 (2003)
  [arXiv:hep-th/0305049].

\bibitem{hep-th/0306018}
  J.~Babington, J.~Erdmenger, N.~J.~Evans, Z.~Guralnik and I.~Kirsch,
  Phys.\ Rev.\ D {\bf 69}, 066007 (2004)
  [arXiv:hep-th/0306018].

\bibitem{hep-th/0311201}
  C.~Nunez, A.~Paredes and A.~V.~Ramallo,
  JHEP {\bf 0312}, 024 (2003)
  [arXiv:hep-th/0311201].

\bibitem{hep-th/0311084}
  P.~Ouyang,
  Nucl.\ Phys.\ B {\bf 699}, 207 (2004)
  [arXiv:hep-th/0311084].

\bibitem{hep-th/0307218}
  X.~J.~Wang and S.~Hu,
  JHEP {\bf 0309}, 017 (2003)
  [arXiv:hep-th/0307218].

\bibitem{hep-th/0312071}
  S.~Hong, S.~Yoon and M.~J.~Strassler,
  JHEP {\bf 0404}, 046 (2004)
  [arXiv:hep-th/0312071].

\bibitem{hep-th/0403279}
  N.~J.~Evans and J.~P.~Shock,
  Phys.\ Rev.\ D {\bf 70}, 046002 (2004)
  [arXiv:hep-th/0403279].

\bibitem{hep-th/0406207}
  B.~A.~Burrington, J.~T.~Liu, L.~A.~Pando Zayas and D.~Vaman,
  JHEP {\bf 0502}, 022 (2005)
  [arXiv:hep-th/0406207].

\bibitem{hep-th/0408113}
  J.~Erdmenger and I.~Kirsch,
  JHEP {\bf 0412}, 025 (2004)
  [arXiv:hep-th/0408113].

\bibitem{hep-th/0410035}
  M.~Kruczenski, L.~A.~P.~Zayas, J.~Sonnenschein and D.~Vaman,
  JHEP {\bf 0506}, 046 (2005)
  [arXiv:hep-th/0410035].

\bibitem{hep-th/0411097}
  S.~Kuperstein,
  JHEP {\bf 0503}, 014 (2005)
  [arXiv:hep-th/0411097].

\bibitem{hep-th/0412260}
  A.~Paredes and P.~Talavera,
  Nucl.\ Phys.\ B {\bf 713}, 438 (2005)
  [arXiv:hep-th/0412260].

\bibitem{hep-th/0412141}
  T.~Sakai and S.~Sugimoto,
  Prog.\ Theor.\ Phys.\ {\bf 113}, 843 (2005)
  [arXiv:hep-th/0412141].

\bibitem{hep-th/0502091}
  N.~Evans, J.~Shock and T.~Waterson,
  JHEP {\bf 0503}, 005 (2005)
  [arXiv:hep-th/0502091].


\bibitem{hep-th/0511044}
  K.~Peeters, J.~Sonnenschein and M.~Zamaklar,
  arXiv:hep-th/0511044.

\bibitem{hep-th/0511045}
  A.~L.~Cotrone, L.~Martucci and W.~Troost,
  arXiv:hep-th/0511045.

\bibitem{hep-ph/0501128}
  J.~Erlich, E.~Katz, D.~T.~Son and M.~A.~Stephanov,
  arXiv:hep-ph/0501128.

\bibitem{hep-ph/0501218}
  L.~Da Rold and A.~Pomarol,
  Nucl.\ Phys.\ B {\bf 721}, 79 (2005)
  [arXiv:hep-ph/0501218].

\bibitem{hep-th/0501022}
  G.~F.~de Teramond and S.~J.~Brodsky,
  Phys.\ Rev.\ Lett.\ {\bf 94}, 201601 (2005)
  [arXiv:hep-th/0501022].

\bibitem{hep-ph/0510334}
  K.~Ghoroku, N.~Maru, M.~Tachibana and M.~Yahiro,
  arXiv:hep-ph/0510334.

\bibitem{hep-ph/0603142}
  J.~Shock and F.~Wu,
  [arXiv:hep-ph/0603142].

\bibitem{hep-ph/0603249}
  N.~Evans and T.~Waterson,
  [arXiv:hep-ph/0603249].

\bibitem{hep-th/0209080}
  H.~Boschi-Filho and N.~R.~F.~Braga,
  Eur.\ Phys.\ J.\ C {\bf 32}, 529 (2004)
  [arXiv:hep-th/0209080].

\bibitem{hep-th/0212207}
  H.~Boschi-Filho and N.~R.~F.~Braga,
  JHEP {\bf 0305}, 009 (2003)
  [arXiv:hep-th/0212207].

\bibitem{hep-th/9906194}
  D.~Z.~Freedman, S.~S.~Gubser, K.~Pilch and N.~P.~Warner,
  JHEP {\bf 0007}, 038 (2000)
  [arXiv:hep-th/9906194].

\bibitem{hep-th/9906201}
  A.~Brandhuber and K.~Sfetsos,
  Adv.\ Theor.\ Math.\ Phys.\  {\bf 3}, 851 (1999)
  [arXiv:hep-th/9906201].

  \bibitem{hep-th/9811167}
  K.~Sfetsos,
  JHEP {\bf 9901}, 015 (1999)
  [arXiv:hep-th/9811167].

\bibitem{hep-th/0510132}
  R.~Hernandez, K.~Sfetsos and D.~Zoakos,
  JHEP {\bf 0603}, 069 (2006)
  [arXiv:hep-th/0510132].

\bibitem{hep-th/0512125}
  A.~Karch, A.~O'Bannon and K.~Skenderis,
  arXiv:hep-th/0512125.

\bibitem{hep-th/0302098}
  D.~E.~Crooks and N.~J.~Evans,
  arXiv:hep-th/0302098.

\bibitem{hep-th/0311270}
  M.~Kruczenski, D.~Mateos, R.~C.~Myers and D.~J.~Winters,
  JHEP {\bf 0405}, 041 (2004)
  [arXiv:hep-th/0311270].

\bibitem{hep-th/0105235}
  J.~Babington, N.~J.~Evans and J.~Hockings,
  JHEP {\bf 0107}, 034 (2001)
  [arXiv:hep-th/0105235].

\end{thebibliography}

\end{document}